\documentstyle[pra,twocolumn,psfig,aps]{revtex} %prl,a,b
\begin{document}
\draft
\title{Recovering Coherence via Conditional Measurements}
\author{M.~Fortunato$^{*(a)}$ \and G.~Harel$^{(b)}$ and G.~Kurizki$^{(b)}$}
\address{(a) Abteilung f\"ur Quantenphysik, Universit\"at Ulm,
          Albert-Einstein-Allee 11, D-89069 Ulm, Germany}
\address{ (b) Dept. of Chemical Physics, The Weizmann Institute of Science,
 Rehovot 76100, Israel}
\date{\today}
\maketitle
\begin{abstract}
We show that conditional measurements on atoms following their 
interaction with a resonant cavity field mode can be used
to effectively counter the decoherence of Fock-state superpositions
due to cavity leakage. 
\end{abstract}
\pacs{PACS numbers: 03.65.Bz, 89.70.+c, 42.50.Dv, 32.80.-t \\% }
%\vskip 4cm
%Corresponding author:\\
%  Gil Harel \\
%  Dept. of Chemical Physics \\ 
%  Weizmann Institute of Science \\
%  Rehovot 76100, Israel \\
%  Tel: 972-8-9343721 \ \ Fax: 972-8-9344123 \\ 
%  E-mail: gil@wiscpa.weizmann.ac.il \\ 
}
%
%\pacs{PACS numbers: 42.50.-p, 42.50.Dv, 32.80.-t, 42.50.Ar}
%03.65.Bz  Foundations, theory of measurement, miscellaneous theories
%          (including Aharonov-Bohm effect, Bell inequalities, Berry's
%          phase)
%32.80.-t  Photon interactions with atoms
%42.50.-p  Quantum optics
%42.50.Ar  Photon statistics and coherence theory
%42.50.Dv  Nonclassical field states; squeezed, antibunched, and
%          sub-Poissonian states; operational definitions of the
%          phase of the field; phase measurements
%42.52.+x  Masers; gyrotrons
%89.70.+c  Information science
%
\section{Introduction}
\label{intro}

Decoherence of non-classical states of a quantum system via coupling to 
a reservoir is of fundamental interest, as it constitutes the mechanism that
yields the classical limit of the system dynamics~\cite{kn:zurek}. 
Recently it has also become a topic of great applied importance, because
it determines the feasibility of quantum information storage, encoding
(encrypting) and computing~\cite{kn:qc}.
In numerous current theoretical proposals, the irreversibility of decoherence 
processes in quantum computing is combatted by two generic means.
One is the filtering out of the ensemble portion which has not decohered, i.e., 
has remained intact. This approach has been suggested for 
two-mode fields~\cite{kn:yama}, but not for single-mode cavity fields.
The other means is encoding the state (qubit) by means of several ancillas, 
decoding the result after a certain time, checking the ancillas for 
error syndromes and correcting them~\cite{kn:others}.
Although the latter approach is in principle applicable to arbitrary errors,
only extremely small error probabilities (per qubit or gate, per time step)
can afford fault tolerant quantum computation \cite{shordoria}.
Instead of the ``high level'' unitary transformation approach to 
error correction in quantum computing---which involves substantial overhead in
qubits and gates---the
countering of decoherence of {\it stored} quantum information,
e.g., in between computation steps, may be achieved by a ``low level'' approach:
applying simple physical manipulations to the quantum storage device,
which take advantage of its specific physical realization. 
Such approach has been advocated recently 
\cite{mabuchi}, relying on continuous monitoring of the dissipation channel
for quantum jumps, with perfect photodetection efficiency, and on instantaneous
feedback for the inversion of their effect.
 
The nature of quantum computing requires that decoherence be corrected without 
knowing which state is in error during the computation. There is, however,
a simpler but still important problem: how to protect from decoherence
the {\it input} states, prior to the onset of computation.
Here we suggest a non-unitary approach to counter decoherence,
which can be used to safely store quantum field states in dissipative cavities,
in order to subsequently use them as {\it apriori known} input in information 
processing or in signal transmission.
The basic idea is to {\it restore} the decohered field state 
by entangling it with an atom, and then projecting the
entangled state onto a superposition of atomic eigenstates,
whose phase and amplitudes are specifically tailored for the field state
we wish to recreate.
Such projection amounts to post-selection of the appropriate atomic state,
i.e., to a conditional measurement (CM)~\cite{kn:cm,kn:ens}.
The specific scheme we put forward is based on modification of our optimized 
CM strategy \cite{kn:gil} for cavity-mode state preparation by resonant
interaction with atoms, in the Jaynes-Cummings (JC) model, 
followed by projection onto selected atomic states.
In the present problem we set the initial (unspoilt) superposition of 
zero-photon up to $N$-photon states as our target state,
and work in Liouville space instead of Hilbert space,
so as to account for the state decoherence.
The results demonstrate that a few {\it highly-probable} CMs, 
in this simple model, can drastically reduce even a large error.
One of our objectives is to find the optimal tradeoff between the CM
probability and the error size, which grows in the course of dissipation. 

The ability to {\it approximately restore any mixture to any pure state} 
(in our $N$+1-dimensional Hilbert space) 
is the advantage of our post-selection CM approach,
compared to the non-selective measurement (tracing) approach:
Mixed states can only evolve into the special ``cotangent" and ``tangent'' 
pure states by a large number of JC interactions with atoms initially prepared 
in superposition states (under the atomic excitation-trapping condition)
followed by tracing over the atomic states~\cite{kn:mey}.

\section{Decoherence Minimization by Conditional Measurements}
\label{dmbcm}

We consider a single-mode cavity in which the quantized electromagnetic field 
is initially prepared in a finite superposition of Fock states, 
\begin{equation}
|\psi(0)\rangle = \sum_{n=0}^N c_n|n\rangle \;.
\label{eq:infieldstate}
\end{equation}
To model the effect of dissipation we assume the
cavity field to be coupled to a zero-temperature heat bath.
The master equation describing such coupling, in the interaction picture, is
\begin{equation}
\dot{\rho}_{_F}  =  \gamma(2\hat{a}\rho_{_F}\hat{a}^\dagger-\hat{a}^\dagger
\hat{a}\rho_{_F}-\rho_{_F}\hat{a}^\dagger\hat{a})\;,
\label{eq:meq}
\end{equation}
where $\rho_{_F}=\rho_{_F}(t)$ is the density matrix of the cavity field,
$\hat{a}$ and $\hat{a}^\dagger$ are the annihilation and creation operators 
of the field, and $\gamma$ is the damping constant of the cavity. 

The solution of Eq.~(\ref{eq:meq}) after dissipation over 
time $\bar{t}>0$ \cite{kn:arnoldus} can be shown to have the form
\begin{eqnarray}
\rho_{n,m}(\bar{t}) & = & \sum_{k=0}^{\infty} 
 \rho_{n+k,m+k}(0)
% \nonumber \\ & & \times 
 \sqrt{{\scriptsize\pmatrix{n+k\cr n} }
 (e^{-2\gamma\bar t})^n 
 (1-e^{-2\gamma\bar t})^k }
 \nonumber \\ & & \times 
 \sqrt{{\scriptsize\pmatrix{m+k\cr m} }
% \sqrt{{\scriptstyle\pmatrix{m+k\cr m} }
 (e^{-2\gamma\bar t})^m 
 (1-e^{-2\gamma\bar t})^k }
 \;,
\label{eq:solro}
\end{eqnarray}
written here in Fock basis,
$\rho_{n,m}(t)=\langle n|\rho_{_F}(t)|m\rangle$. 

In order to recover the original state of the field
we propose to apply 
an optimized CM (or a sequence thereof) to the cavity as follows: 
Using a classical field we prepare a two-level atom in a chosen 
superposition~\cite{kn:gil,kn:meschede}
\begin{equation}
|\phi^{(i)}\rangle = \alpha^{(i)} |e\rangle + \beta^{(i)}|g\rangle
\label{eq:inatstate}
\end{equation}
of its ground $|g\rangle$ and excited $|e\rangle$ states, 
and let it interact with the field for a time $\tau$ by sending it through the 
cavity with controlled speed.
The field-atom interaction is adequately described by the resonant 
Jaynes-Cummings (JC) model~\cite{kn:jc}.
We assume the field-atom interaction time $\tau$ to be much shorter than the
cavity lifetime, $\gamma\tau \ll 1$,
so that we may neglect dissipation during each CM.
Upon exiting the cavity the atom is {\it conditionally measured},
using a second classical field, to be in a state 
\begin{equation}
|\phi^{(f)}\rangle = \alpha^{(f)} |e\rangle +
                            \beta^{(f)}|g\rangle\;,
\label{eq:finatstate}
\end{equation}
differing in general from the initial atomic state $|\phi^{(i)}\rangle$. 
This means that we post-select, using the same setup as in ref.~\cite{kn:gil},
the atomic superposition state~(\ref{eq:finatstate}) which is {\it correlated}
to a cavity field state that is as close as possible to the original 
state~(\ref{eq:infdm}).
%
%\section{Results on the density matrix}
\label{dynamics}

The effect the applied CM has on the cavity field is then calculated as follows:
Initially, at the time the atom enters the cavity, 
the density matrix of the field-atom system is 
\begin{equation}
\rho_{_{FA}}(\bar{t})=
 \rho_{_F}(\bar{t}) \otimes |\phi^{(i)}\rangle\langle\phi^{(i)}|\;.
\label{eq:intotdens}
\end{equation}
It then evolves unitarily by the JC interaction of duration $\tau$ into 
\begin{equation}
\rho_{_{FA}}(\bar{t}+\tau)=\hat{U}(\tau)
\rho_{_{FA}}(\bar{t})\hat{U}^\dagger(\tau)\;,
\label{eq:evolvrho}
\end{equation}
where $\hat{U}(\tau)$ is the interaction picture evolution operator
\begin{eqnarray}
\hat{U}(\tau) |n\rangle|e\rangle & = & 
 C_n |n\rangle|e\rangle -iS_n |n+1\rangle|g\rangle 
\label{eq:uneva} \nonumber\\
\hat{U}(\tau) |n\rangle|g\rangle & = & 
 C_{n-1} |n\rangle|g\rangle -iS_{n-1}|n-1\rangle|e\rangle\; ,
\label{eq:unevb}
\end{eqnarray}
with $C_n=\cos\left(\lambda\tau\sqrt{n+1}\right)$ and 
$S_n=\sin\left(\lambda\tau\sqrt{n+1}\right)$,
$\lambda$ being the field-atom coupling constant (known as the vacuum Rabi 
frequency).
Finally, the conditional measurement of the atom in the state 
$|\phi^{(f)}\rangle$ results in a density matrix of the {\it field} given by 
\begin{equation}
\rho_{_F}(\bar{t}+\tau)={\rm Tr}_{_A} \left[\rho_{_{FA}}(\bar{t}+\tau)
|\phi^{(f)}\rangle\langle\phi^{(f)}|\right]/P\;,
\label{eq:traceformula}
\end{equation}
where
\begin{equation}
P={{\rm Tr}_{_{F}}{\rm Tr}_{_A}\left[\rho_{_{FA}}(\bar{t}+\tau)
|\phi^{(f)}\rangle\langle\phi^{(f)}|\right]}
\label{eq:probability}
\end{equation}
is the success probability of the CM. 
The explicit expressions for $\rho_{_F}(\bar{t}+\tau)$ and $P$ are given in 
the Appendix for an initial superposition of $|0\rangle$ and $|1\rangle$ states.

To nearly recover the original state of the field, we use the dependence of 
$\rho_{_{F}}(\bar{t}+\tau)$ on the initial and final atomic states and the 
field-atom interaction time,
choosing {\it optimal} parameters $\alpha^{(i)}$, $\beta^{(i)}$, 
$\alpha^{(f)}$, $\beta^{(f)}$ and $\tau$ such that
\begin{equation}
\rho_{_{F}}(\bar{t}+\tau) \approx \rho_{_{F}}(0) 
\label{rc}
\end{equation}
holds (see Appendix for an explicit form of this condition), 
along with high CM success probability  (\ref{eq:probability}).
These optimal CM parameters are found by minimizing the cost 
function~\cite{kn:gil} 
\begin{equation}
G=\frac{d(\rho_{_{F}}(\bar{t}+\tau),\rho_{_{F}}(0))}{P^r}\;, 
\label{eq:cf} \end{equation}
where $d$ is a distance function between two density matrices, defined as
\begin{equation} d(\rho_{_F}^{(1)},\rho_{_F}^{(2)})=
 \sqrt{\sum_{nm}(\rho^{(1)}_{nm}-\rho^{(2)}_{nm})^2}\;,
\label{eq:df}\end{equation}
$P$ is the CM success probability (\ref{eq:probability}),
and the adjustable exponent $r>0$ determines the relative importance of the two 
factors in $G$.
If this CM does not bring us as close to the original state as our experimental
accuracy permits, we can repeat the process over and over again, as long
as the distance to the original state keeps decreasing, while the CM success
probability remains high.  The atomic states (\ref{eq:inatstate}) and 
(\ref{eq:finatstate}) are determined by the {\it minimization} of 
(\ref{eq:cf}) at each step.
Let us note here that the application of each CM may introduce
widening of the photon-number distribution by one photon,
and yet the {\it optimized} CMs are capable of avoiding this widening and,
moreover, of restoring the field to its initial pure state.
Eqs. (\ref{eq:eqe}-\ref{eq:eqi}) in the Appendix exemplify the 
widening-avoidance requirements which are implicit in condition (\ref{rc}).
These requirements amount to an effective control of a large Fock-state 
subspace.

\section{Examples}
\label{examples}

We illustrate our approach with two examples below, using the $Q$-function
$ Q_{\rho_{_F}}(\alpha,\alpha^*)= \langle\alpha|\rho_{_F}|\alpha\rangle$,
$|\alpha\rangle$ being a coherent state of complex amplitude $\alpha$, 
to visualize the error-correction process:

1) Let us take as the original field state an equal-amplitude
superposition of our basis states, {\it e.g.}, 
\begin{equation}
|\psi(0)\rangle=(|0\rangle+e^{i\pi/3}|1\rangle)/\sqrt2\;,
\end{equation}
whose $Q$-function is shown in Fig. 1(a).
Dissipation by $\gamma \bar{t}=0.3$ renders the {\it error} matrix
$ \rho_{_F}(\bar t)-\rho_{_F}(0) $ of considerable magnitude,
as seen in Fig. 1(b).
After the application of one CM
($|\phi^{(i)}\rangle =\cos(3\pi/8)|e\rangle + \sin(3\pi/8)e^{i5\pi/4}|g\rangle$,
 $\lambda\tau=37.95$,
 $|\phi^{(f)}\rangle =\cos(3\pi/8)|e\rangle + \sin(3\pi/8)e^{i\pi/4}|g\rangle$),
 optimized to yield high success probability
($r=2$), the remaining error matrix 
$ \rho_{_F}(\bar t+\tau)-\rho_{_F}(0) $ is roughly 2.5 times smaller than 
before the correction, as seen in Fig. 1(c). 
The success probability of the CM is a high 74\%.
Subsequent CMs can further reduce the distance to $1/6$ (one sixth) its original
magnitude, with 62\% success probability for the full CM sequence (Fig. 3).
Stronger error reduction is obtainable at the expense of success probability: 
the application of 4 CMs optimized for $r=1$ (respectively $r=0$) yields an 
error reduction 
factor of 11 (respectively 28) with sequence probability of 
33\% (respectively 16\%).
%1111111111111111111111111111111111111111111111111111111111111
%%%%%%%%%%%%%%%%%%%%%%%%%%%%%%%%%%%%%%%%%%%%%%%%%%%%%%%%%%%%%%

2) If the original field state is a strongly unequal superposition of the
basis states, such as
\begin{equation}
|\psi(0)\rangle=10^{-1}|0\rangle+e^{i\pi/3}\sqrt{1-10^{-2}}|1\rangle
\label{eq:infdmex2}
\end{equation}
(Fig. 2(a)), the error matrix after dissipation by $\gamma \bar{t}=0.3$
is again significant (Fig. 2(b)).
Successive application of 4 CMs, optimized for $r=2$,
reduces this error by a factor of 30 (Fig. 2(c)), which means that
the recovered state is {\it practically indistinguishable} from the
original state.
The success probability of the {\it total} CM sequence, 50\%, is 
markedly high (Fig. 3).
If we ignore success probability in Eq. (\ref{eq:cf}) ($r=0$)
we obtain a higher error-reduction factor of 75, with sequence probability
of 28\%.
%%\vskip -.1 cm
%2222222222222222222222222222222222222222222222222222222222222
%%%%%%%%%%%%%%%%%%%%%%%%%%%%%%%%%%%%%%%%%%%%%%%%%%%%%%%%%%%%%%

In Fig. 3 we plot the distance $d_K=d(\rho^K,\rho_{_F}(0))$ (Eq. (\ref{eq:df}))
between the recovered state and the original state
and the CM sequence probability $P_{seq,K}=\prod_{l=1}^K P_l$,
with
$P_l$ given by~(\ref{eq:probability}), as a function
of the number of CMs performed. It shows that the first CMs achieve
a strong reduction of such a distance, whereas after a few successive CMs
saturation sets on, in terms of both distance and success probability.
%due to the high value chosen for $r$ in~(\ref{eq:cf}).
%3333333333333333333333333333333333333333333333333333333333333
%%%%%%%%%%%%%%%%%%%%%%%%%%%%%%%%%%%%%%%%%%%%%%%%%%%%%%%%%%%%%%

It is interesting to compare the success probability in our 
approach with the {\it theoretical} probability to find the original state
in the dissipation-spoilt state,
namely, ${\rm Tr}_{_F}[\rho_{_F}(0)\rho_{_F}(\bar t)]$, which we call
the {\it filtering probability}.
In Table I we list the success probability of a sequence of 4 CMs 
(optimized for $r=2$), 
$P_{seq,K=4}$, and the corresponding filtering probability 
for various values of the dissipation parameter $\gamma\bar t$, 
taking as the original state the state (\ref{eq:infdmex2}) of example 2.
The probability ${\rm Tr}_{_F}[\rho_{_F}(0)\rho_{_F}^{K=4}]$ of finding the 
original state in the recovered state is 0.99 or higher
for all entries.
\section{Discussion}
\label{conclu}

In conclusion, we have demonstrated here the effectiveness of
simple JC-dynamics CMs as a means of reversing the effect of dissipation 
on coherent superpositions of Fock-states of a cavity field: 
the application of a small number of optimized CMs 
recovers the original state of the field with high success probability,
which {\it is comparable or even surpasses} the filtering probability. 
The simplest tactics may employ a {\it single highly-probable trial} 
to achieve nearly-complete error correction.
As noted above, although we have only five control parameters at our
disposal for each CM, our optimization procedure is able to effectively 
control the amplitudes in a large Fock-state subspace.

Among the experimental imperfections that can degrade the effectiveness
of any CM approach~\cite{kn:gil,kn:kozh,kn:gil2},
realistic atomic velocity fluctuations (of 1\%) and cavity-temperature
effects (below $1^\circ$K) are relatively unimportant,
and especially so in the present scheme which makes use of a single or few CMs
so that the effect of experimental imperfections is linear in the input errors.
Only atomic detection efficiency is an experimental challenge~\cite{kn:gil2}.
Although the detection efficiency is currently low, it is expected to rise
considerably in the coming future.

Extensions of this approach to field-atom interaction Hamiltonians with more
controllable degrees of freedom~\cite{kn:kozh} can make a single trial within 
this correction procedure effective for highly complicated states, 
encoding many qubits of information.
Nevertheless, even in its present form the suggested approach has undoubted
merits:
(a) it can yield higher success probabilities than the filtering approach;
(b) it is not limited to small errors as ``high level'' unitary-transformation
approaches are;
(c) it corrects errors after their occurrence,
with no reliance on ideal continuous monitoring of the dissipation
channel and on instantaneous feedback; and
(d) it is realistic in that it can counter combined phase-amplitude errors
which arise in cavity dissipation, and is of general applicability---not
restricted to specific models of dissipation.
\acknowledgments
\label{acknow}

The support of the German-Israeli Foundation (GIF) is acknowledged.
M.F. thanks the European Economic Community
(Human Capital and Mobility programme) for support.
\appendix
\section*{}
The reduced density matrix of the field resulting from its interaction with
the atom followed by the conditional measurement on the latter can be found
using the formula (Eqs. (\ref{eq:intotdens}-\ref{eq:probability}))
\begin{eqnarray}
\rho_{_F}&(&\bar{t}+\tau)= 
 \nonumber\\ && 
{\rm Tr}_{_A}\left[
\hat{U}(\tau)
\rho_{_F}(\bar{t})\!\otimes\!|\phi^{(i)}\rangle\langle\phi^{(i)}|
\hat{U}^\dagger(\tau)
|\phi^{(f)}\rangle\langle\phi^{(f)}|
\right]/P\;,
 \nonumber\\  
\label{eq:traceformula2}
\end{eqnarray}
where the normalization constant $P$ is the success probability of the
conditional measurement and is given by
\begin{eqnarray}
P = && {\rm Tr}_{_F}{\rm Tr}_{_A}\left[
\hat{U}(\tau)
\rho_{_F}(\bar{t})\!\otimes\!|\phi^{(i)}\rangle\langle\phi^{(i)}|
\hat{U}^\dagger(\tau)
|\phi^{(f)}\rangle\langle\phi^{(f)}|
\right]\;.
 \nonumber\\
\label{eq:probability2}
\end{eqnarray}
In the simple case where the initial field state is a superposition of the 
vacuum and one-photon states
\begin{equation}
|\psi(0)\rangle = c_0|0\rangle + c_1|1\rangle
 \;,\;\;\;\rho_{_F}(0)=|\psi(0)\rangle\langle\psi(0)| \;,
\label{eq:infdm}
\end{equation}
the density matrix resulting from dissipation over time 
$\bar{t}$ is
\begin{eqnarray}
\rho_{_F}(\bar{t}) & = & \rho_{00}(\bar{t})|0\rangle\langle 0| 
            + \rho_{01}(\bar{t})|0\rangle\langle 1| +
\nonumber \\
            & & \; + \; \rho_{10}(\bar{t})|1\rangle\langle 0|
            +\rho_{11}(\bar{t})|1\rangle\langle 1| \;,
\label{eq:rhot}
\end{eqnarray}
with
\begin{eqnarray}
\rho_{00}(\bar{t}) & = & |c_0|^2+(1-e^{-2\gamma\bar t})|c_1|^2 \nonumber \\
\rho_{01}(\bar{t}) & = & e^{-\gamma\bar t}c_0c_1^* \nonumber \\
\rho_{10}(\bar{t}) & = & e^{-\gamma\bar t}c_1c_0^* \nonumber  \\
\rho_{11}(\bar{t}) & = & e^{-2\gamma\bar t}|c_1|^2 \;. 
\label{eq:rho}
\end{eqnarray}
The explicit expressions for $\rho_{_F}(\bar{t}+\tau)$ and the 
success probability $P$ are then
\begin{eqnarray}
&\rho_{_F}&(\bar{t}+\tau)=P^{-1}\Big\{\left[|\alpha^{({\rm f})}
|^2A+\alpha^{({\rm f})}\beta^{{({\rm f})}*}M+\beta^{({\rm f})}
\alpha^{({\rm f})*}C\right.
\nonumber \\
&+& \left.\!|\beta^{({\rm f})}|^2O\right]|0\rangle\langle 0|
    +\!\left[|\alpha^{({\rm f})}|^2B +\alpha^{({\rm f})}\beta^{{({\rm f})}*}N
    +\beta^{({\rm f})}\alpha^{({\rm f})*}D\right.
\nonumber \\
& + & \left.\!|\beta^{({\rm f})}|^2K\right]
    |0\rangle\langle 1|+\!\left[|\alpha^{({\rm f})}|^2E
    +\alpha^{({\rm f})}\beta^{{({\rm f})}*}R
    +\beta^{({\rm f})}\alpha^{({\rm f})*}H\right.
\nonumber \\
& + & \left.\!|\beta^{({\rm f})}|^2T\right]|1\rangle\langle 0|+
      \!\left[|\alpha^{({\rm f})}|^2F+\alpha^{({\rm f})}\beta^{{({\rm f})}*}S
      +\beta^{({\rm f})}\alpha^{({\rm f})*}I\right.
\nonumber \\
& + & \left.\!|\beta^{({\rm f})}|^2U\right]|1\rangle\langle 1|+\!\left[
    \alpha^{({\rm f})}\beta^{{({\rm f})}*}W+|\beta^{({\rm f})}|^2Y\right]
    |2\rangle\langle 0|
\nonumber \\
& + & \!\left[\beta^{{({\rm f})}}\alpha^{({\rm f})*}G+|\beta^{({\rm f})}|^2Q
      \right]|0\rangle\langle 2|+\!\left[\alpha^{({\rm f})}\beta^{{({\rm f})}*}
      X\right.
\nonumber \\
& + & \left.\!|\beta^{({\rm f})}|^2J\right]|2\rangle\langle 1|+
      \!\left[\beta^{{({\rm f})}}\alpha^{({\rm f})*}L+|\beta^{({\rm f})}|^2V
      \right]|1\rangle\langle 2|
\nonumber \\
& + & |\beta^{({\rm f})}|^2Z\;|2\rangle\langle 2|\Big\}\;,
\label{eq:findens}
\end{eqnarray}
and
\begin{eqnarray}
P &=&|\alpha^{({\rm f})}|^2(A+F)+\alpha^{({\rm f})}\beta^{{({\rm f})}*}
        (M+S)
\nonumber \\
      & &\; +\; \beta^{({\rm f})}\alpha^{({\rm f})*}(C+I)+|\beta^{({\rm f})}|^2
        (O+U+Z)\;.
\label{eq:norm}
\end{eqnarray}
The coefficients $A,B,\ldots$ here are given by
\begin{mathletters}
\begin{eqnarray}
A&=&\rho_{00}(\bar{t})|\alpha^{({\rm i})}|^2C_0^2
    +i\rho_{01}(\bar{t})\alpha^{({\rm i})}\beta^{({\rm i})*}C_0S_0 \nonumber \\
&&\;+\;\rho_{11}(\bar{t})|\beta^{({\rm i})}|^2S_0^2
  -i\rho_{10}(\bar{t})\alpha^{({\rm i})*}\beta^{({\rm i})}S_0C_0\;,
\label{eq:coeffa} \\
B&=&\rho_{01}(\bar{t})|\alpha^{({\rm i})}|^2C_0C_1-i\rho_{11}(\bar{t})
    \alpha^{({\rm i})*}\beta^{({\rm i})}C_1S_0\;,
\label{eq:coeffb} \\
C&=&\rho_{00}(\bar{t})\alpha^{({\rm i})}\beta^{({\rm i})*}C_0
    -i\rho_{10}(\bar{t})|\beta^{({\rm i})}|^2S_0\;,
\label{eq:coeffc} \\
D&=&i\rho_{00}(\bar{t})|\alpha^{({\rm i})}|^2C_0S_0-i\rho_{11}(\bar{t})
    |\beta^{({\rm i})}|^2S_0C_0 \nonumber \\
&&\;+\;\rho_{01}(\bar{t})\alpha^{({\rm i})}\beta^{({\rm i})*}C_0^2
  +\rho_{10}(\bar{t})\alpha^{({\rm i})*}\beta^{({\rm i})}S_0^2\;,
\label{eq:coeffd} \\
F&=&\rho_{11}(\bar{t})|\alpha^{({\rm i})}|^2C_1^2\;,
\label{eq:coefff} \\
G&=&\rho_{11}(\bar{t})\alpha^{({\rm i})*}\beta^{({\rm i})}S_0S_1
    +i\rho_{01}(\bar{t})|\alpha^{({\rm i})}|^2C_0S_1\;,
\label{eq:coeffg} \\
H&=&\rho_{10}(\bar{t})\alpha^{({\rm i})}\beta^{({\rm i})*}C_1\;,
\label{eq:coeffh} \\
I&=&\rho_{11}(\bar{t})\alpha^{({\rm i})}\beta^{({\rm i})*}C_1C_0
    +i\rho_{10}(\bar{t})|\alpha^{({\rm i})}|^2C_1S_0\;,
\label{eq:coeffi} \\
K&=&i\rho_{00}(\bar{t})\alpha^{({\rm i})*}\beta^{({\rm i})}S_0
    +\rho_{01}(\bar{t})|\beta^{({\rm i})}|^2C_0\;,
\label{eq:coeffp} \\
L&=&i\rho_{11}(\bar{t})|\alpha^{({\rm i})}|^2C_1S_1\;,
\label{eq:coeffl} \\
O&=&\rho_{00}(\bar{t})|\beta^{({\rm i})}|^2\;,
\label{eq:coeffo} \\
Q&=&i\rho_{01}(\bar{t})\alpha^{({\rm i})*}\beta^{({\rm i})}S_1\;,
\label{eq:coeffq}\\
U&=&\rho_{00}(\bar{t})|\alpha^{({\rm i})}|^2S_0^2
    -i\rho_{01}(\bar{t})\alpha^{({\rm i})}\beta^{({\rm i})*}S_0C_0 \nonumber \\
&&\;+\;\rho_{11}(\bar{t})|\beta^{({\rm i})}|^2C_0^2
  +i\rho_{10}(\bar{t})\alpha^{({\rm i})*}\beta^{({\rm i})}C_0S_0\;,
\label{eq:coeffu} \\
V&=&i\rho_{11}(\bar{t})\alpha^{({\rm i})*}\beta^{({\rm i})}C_0S_1
    +\rho_{01}(\bar{t})|\alpha^{({\rm i})}|^2S_0S_1\;,
\label{eq:coeffv} \\
Z&=&\rho_{11}(\bar{t})|\alpha^{({\rm i})}|^2S_1^2\;,
\label{eq:coeffz}
\end{eqnarray}
\label{eq:coeff}
\end{mathletters}
with the following relations holding between them
\begin{mathletters}
\begin{eqnarray}
A&=&A^*\;,\;\;\;B=E^*\;,\;\;\;C=M^*\;,\;\;\;D=R^* \label{eq:rela} \\
F&=&F^*\;,\;\;\;G=W^*\;,\;\;\;H=N^*\;,\;\;\;I=S^* \label{eq:relb} \\
L&=&X^*\;,\;\;\;O=O^*\;,\;\;\;K=T^*\;,\;\;\;Q=Y^* \label{eq:relc} \\
U&=&U^*\;,\;\;\;V=J^*\;,\;\;\;Z=Z^*\;. \label{eq:reld}
\end{eqnarray}
\label{eq:rel}
\end{mathletters}
\vskip -.3cm \noindent
(The coefficients $K$ and $N$ in this appendix bear no relation to
$K$ and $N$ mentioned in the main text).

The explicit form of condition (\ref{rc}) for recovering the original 
field state is given by the following list of approximation relations:
\begin{mathletters}
\begin{eqnarray}
\rho_{00}(\bar{t}+\tau)&=&
P^{-1}\left[|\alpha^{({\rm f})}|^2A+\alpha^{({\rm f})}
\beta^{{({\rm f})}*}M
\right.  \nonumber \\ & & \left.\; + \; 
\beta^{({\rm f})}\alpha^{({\rm f})*}C
+|\beta^{({\rm f})}|^2O\right] \approx |c_0|^2
\label{eq:eqa} \\
\rho_{01}(\bar{t}+\tau)&=&
P^{-1}\left[|\alpha^{({\rm f})}|^2B
+\alpha^{({\rm f})}\beta^{{({\rm f})}*}N
\right.  \nonumber \\ & & \left.\; + \; 
\beta^{({\rm f})}\alpha^{({\rm f})*}D
+|\beta^{({\rm f})}|^2K\right]
\approx c_0c_1^*
\label{eq:eqb} \\
\rho_{10}(\bar{t}+\tau)&=&
P^{-1}\left[|\alpha^{({\rm f})}|^2E
+\alpha^{({\rm f})}\beta^{{({\rm f})}*}R
\right.  \nonumber \\ & & \left.\; + \; 
\beta^{({\rm f})}\alpha^{({\rm f})*}H
+|\beta^{({\rm f})}|^2T\right]
\approx c_1c_0^*
\label{eq:eqc} \\
\rho_{11}(\bar{t}+\tau)&=&
P^{-1}\left[|\alpha^{({\rm f})}|^2F+\alpha^{({\rm f})}
\beta^{{({\rm f})}*}S
\right.  \nonumber \\ & & \left.\; + \; 
\beta^{({\rm f})}\alpha^{({\rm f})*}I
+|\beta^{({\rm f})}|^2U\right]
\approx |c_1|^2
\label{eq:eqd} \\
\rho_{02}(\bar{t}+\tau)&=&
P^{-1}\left[\beta^{{({\rm f})}}\alpha^{({\rm f})*}G+|\beta^{({\rm f})}|^2Q
\right] \approx 0
\label{eq:eqe}\\
\rho_{20}(\bar{t}+\tau)&=&
P^{-1}\left[\alpha^{({\rm f})}\beta^{{({\rm f})}*}W+|\beta^{({\rm f})}|^2Y 
\right] \approx 0
\label{eq:eqf}\\
\rho_{12}(\bar{t}+\tau)&=&
P^{-1}\left[\beta^{{({\rm f})}}\alpha^{({\rm f})*}L+|\beta^{({\rm f})}|^2V 
\right] \approx 0
\label{eq:eqg} \\
\rho_{21}(\bar{t}+\tau)&=&
P^{-1}\left[\alpha^{({\rm f})}\beta^{{({\rm f})}*}X+|\beta^{({\rm f})}|^2J 
\right] \approx 0
\label{eq:eqh} \\
\rho_{22}(\bar{t}+\tau)&=& 
P^{-1}\left[|\beta^{({\rm f})}|^2Z \right] \approx 0 \;.
\label{eq:eqi}
\end{eqnarray}
\label{eq:eq}
\end{mathletters}
%

%1111111111111111111111111111111111111111111111111111111111111
\begin{figure}
%\begin{picture}(0,0)
%\put(-00,-35){\makebox(0,0)[c]{(a)}}
%\put(-00,-205){\makebox(0,0)[c]{(b)}}
%\put(-00,-385){\makebox(0,0)[c]{(c)}}
%\end{picture}
{\centerline{\vbox{
\psfig{file=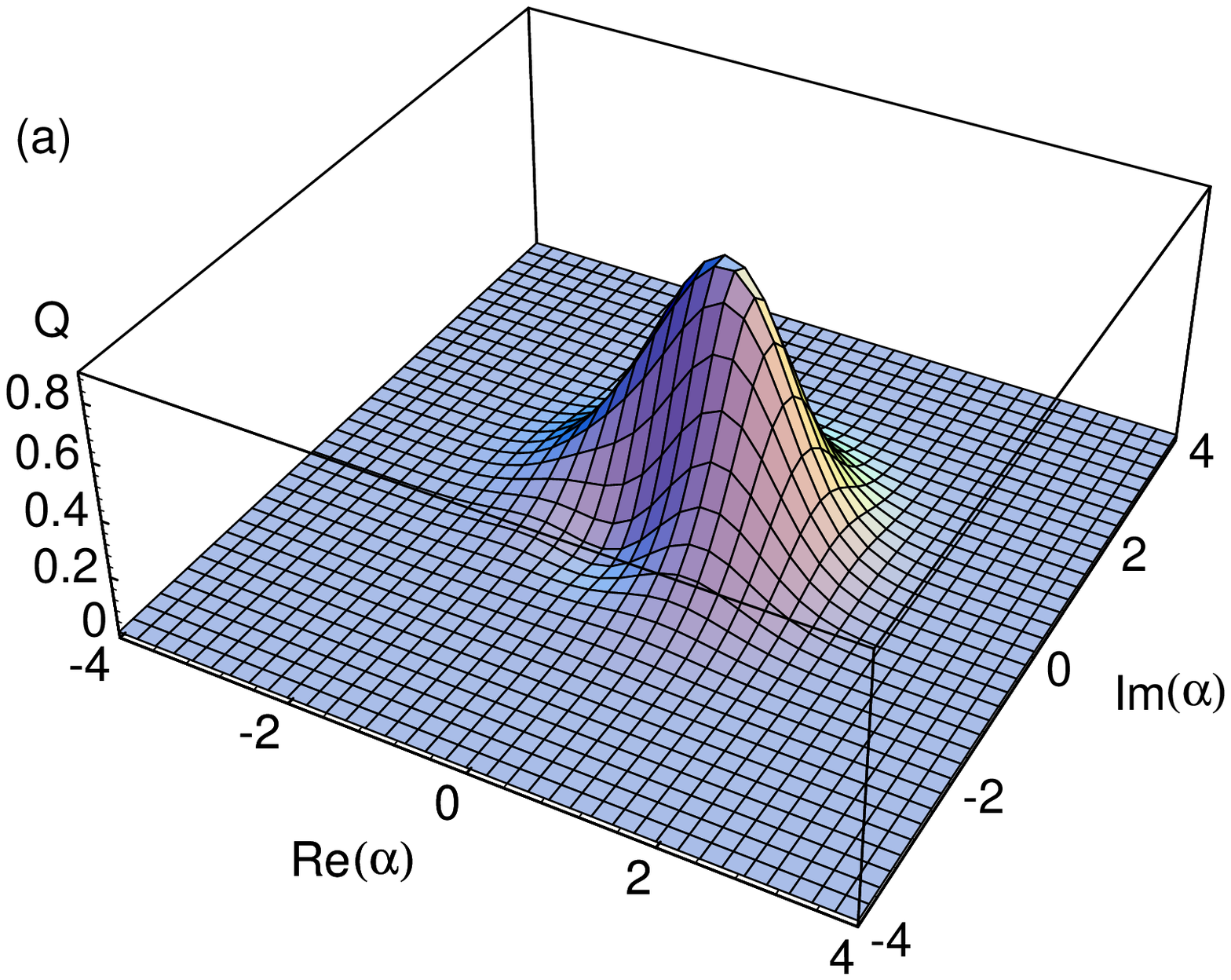,width=7.5cm,angle=00}
\psfig{file=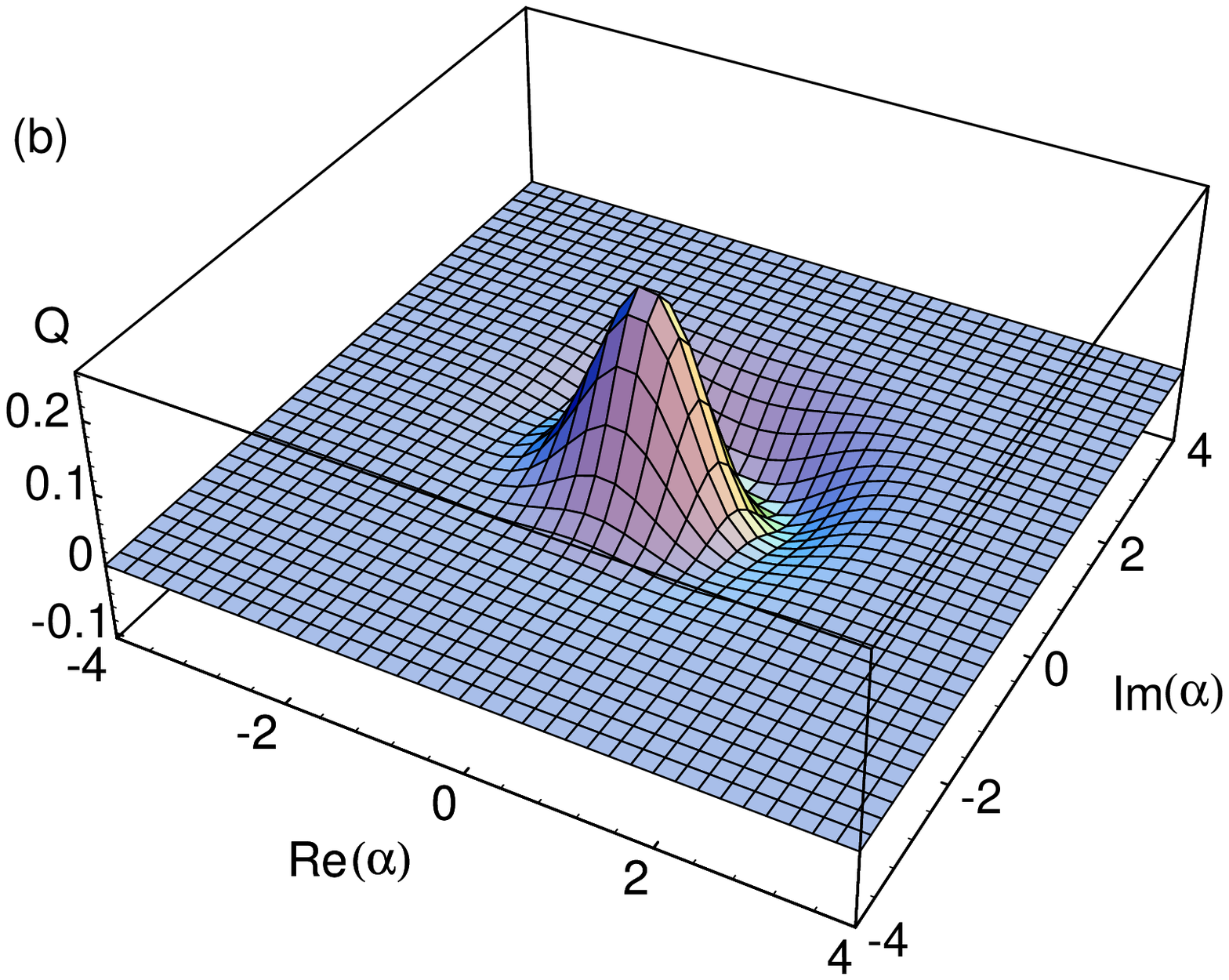,width=7.5cm,angle=00}
\psfig{file=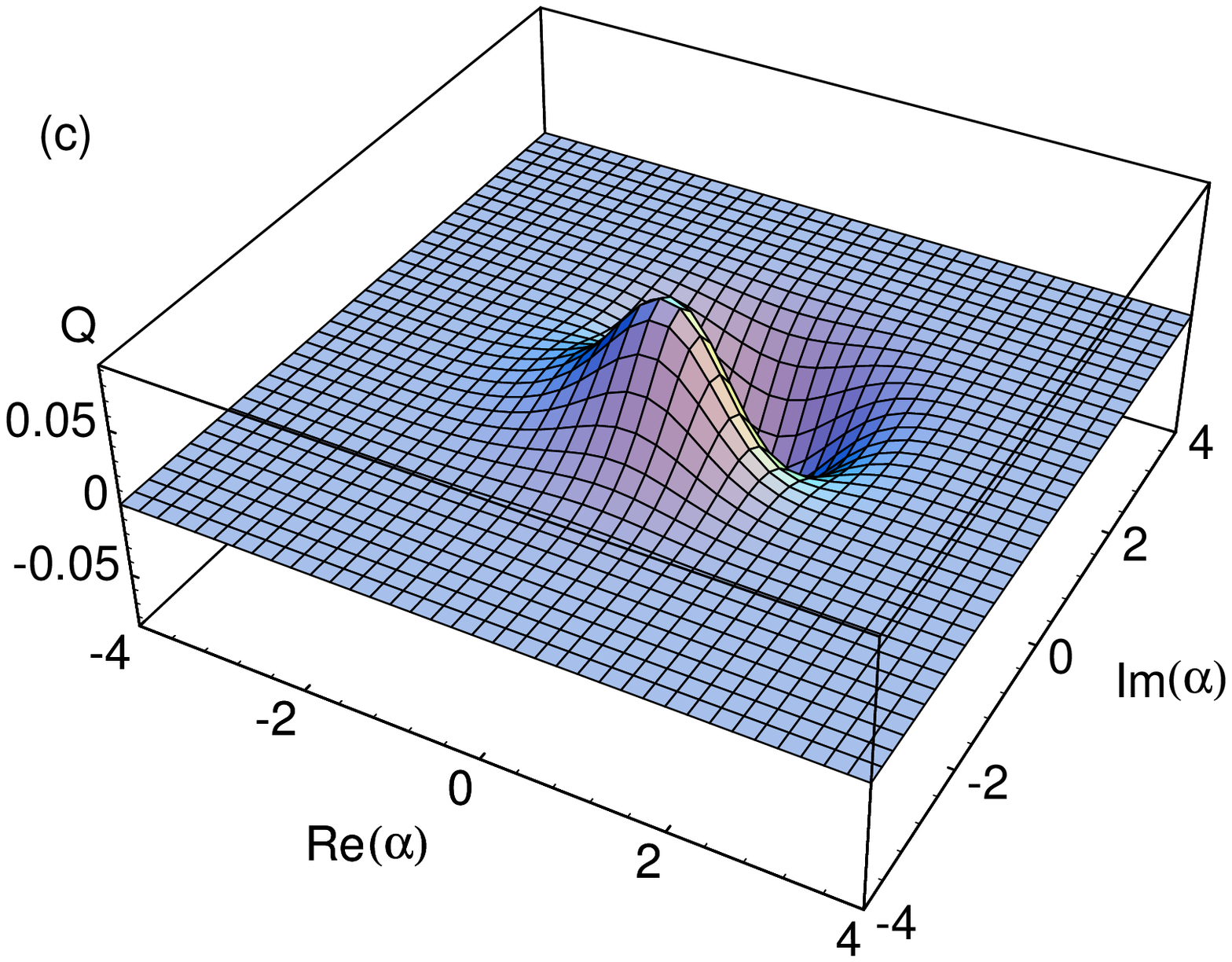,width=7.5cm,angle=00}
}}}
\vskip .3 cm
\caption{
$Q$-function description of
(a) original field, with $\rho_{_F}(0)=|\psi(0)\rangle\langle\psi(0)|$,
$|\psi(0)\rangle=(|0\rangle+e^{i\pi/3}|1\rangle)/\protect\sqrt2$;
(b) error after dissipation, $\rho_{_F}(\bar t)-\rho_{_F}(0)$;
(c) reduced error after 1 optimized CM (minimizing Eq. (\protect\ref{eq:cf})),
 $\rho_{_F}(\bar t+\tau)-\rho_{_F}(0)$.
}
\end{figure}
%2222222222222222222222222222222222222222222222222222222222222
\begin{figure}
%\begin{picture}(0,0)
%\put(-00,-35){\makebox(0,0)[c]{(a)}}
%\put(-00,-205){\makebox(0,0)[c]{(b)}}
%\put(-00,-385){\makebox(0,0)[c]{(c)}}
%\end{picture}
{\centerline{\vbox{
\psfig{file=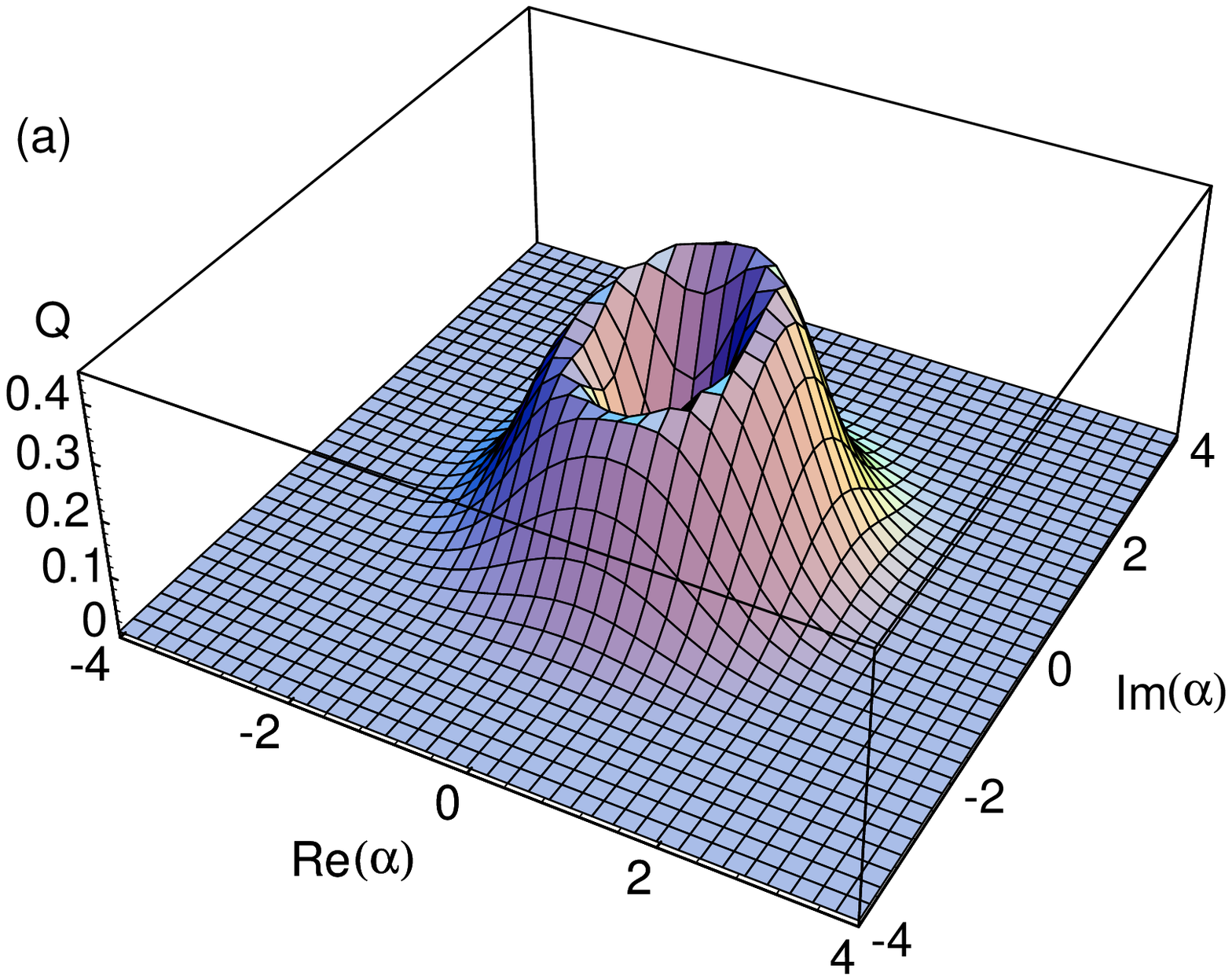,width=7.5cm,angle=00}
\psfig{file=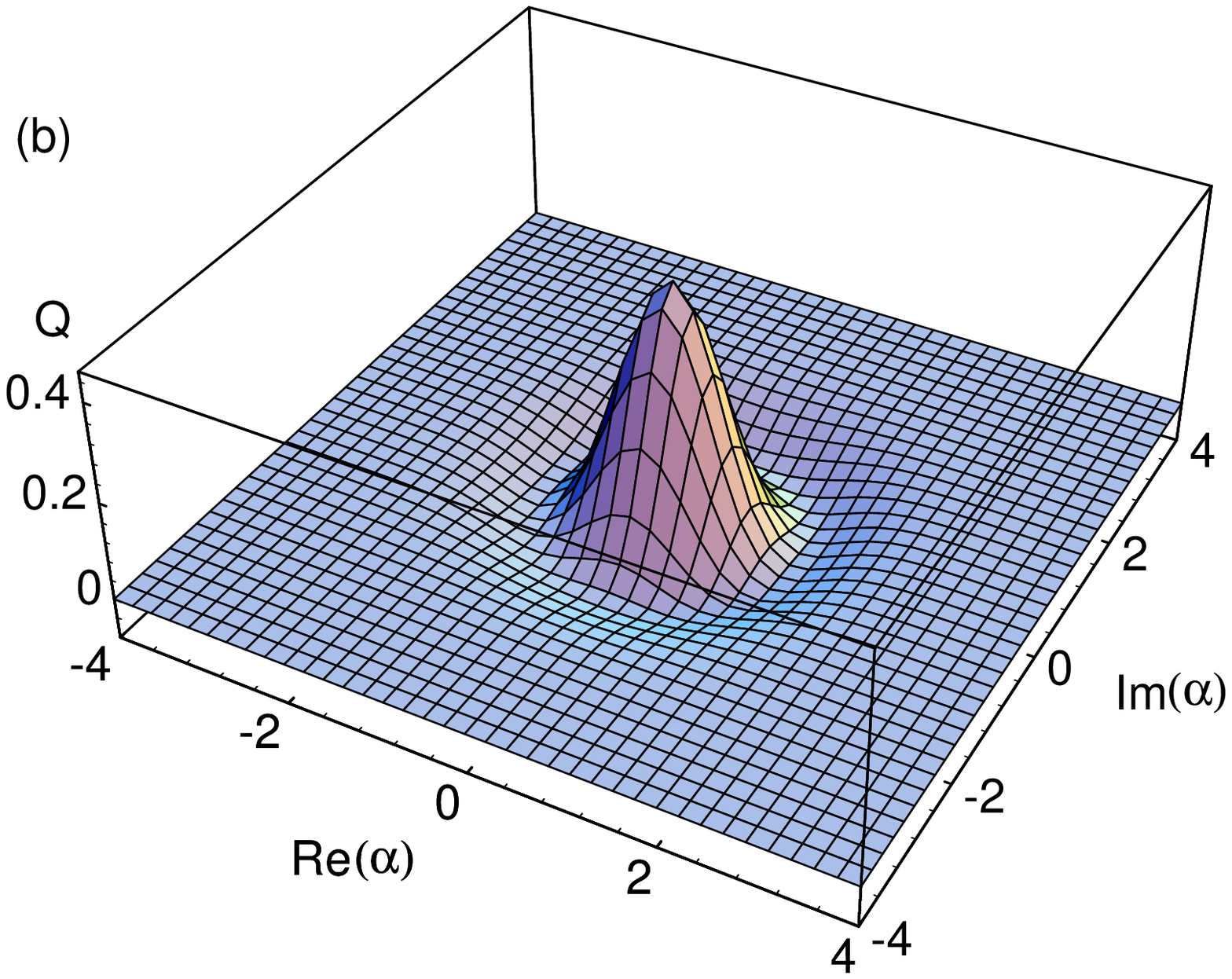,width=7.5cm,angle=00}
\psfig{file=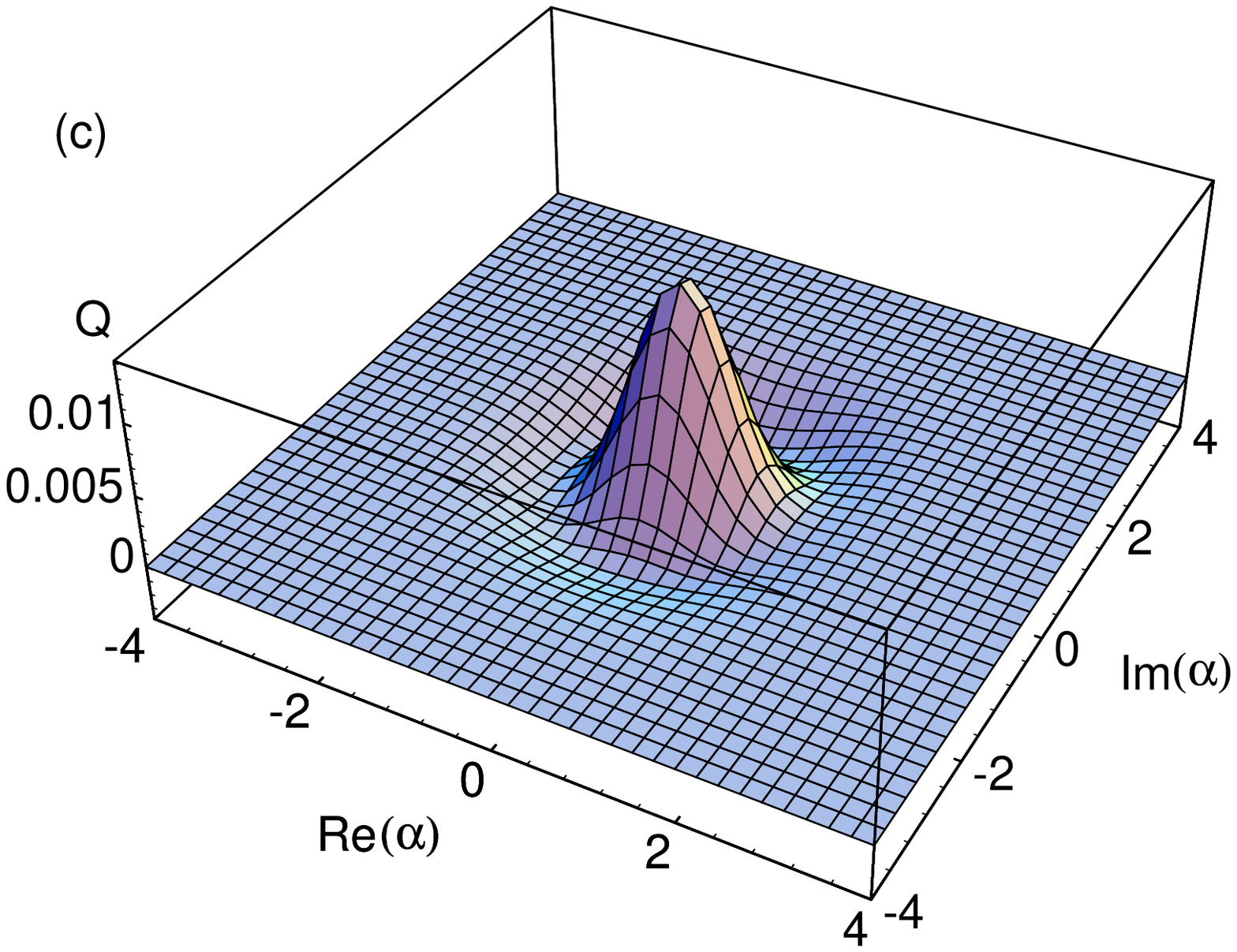,width=7.5cm,angle=00}
}}}
\vskip .3 cm
\caption{
$Q$-function description of 
(a) original field,
$\rho_{_F}(0)=|\psi(0)\rangle\langle\psi(0)|$,
$|\psi(0)\rangle=10^{-1}|0\rangle+e^{i\pi/3}\protect\sqrt{1-10^{-2}}|1\rangle$;
(b) error after dissipation, $\rho_{_F}(\bar t)-\rho_{_F}(0)$;
(c) error after 4 consecutive optimized CMs, each minimizing Eq. 
 (\protect\ref{eq:cf}).
}
\end{figure}
%3333333333333333333333333333333333333333333333333333333333333
\begin{figure}
{\centerline{\vbox{
\psfig{file=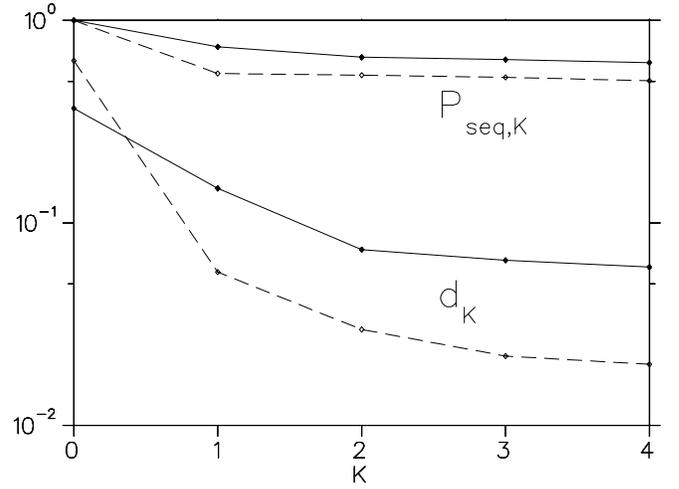,width=8.6cm,angle=90}
}}}
\vskip .3 cm
\caption{Distance $d_K$ and probability $P_{seq,K}$ as functions of 
the number $K$ of CMs applied to the cavity. 
Solid curves -- example 1; dashed curves -- example 2. 
}
\end{figure}
%%%%%%%%%%%%%%%%%%%%%%%%%%%%%%%%%%%%%%%%%%%%%%%%%%%%%%%%%%%%%%

%%%%%%%%%%%%%%%%%%%%%%%%%%%%%%%%%%%%%%%%%%%%%%%%%%%%%%%%%%%%%%
\begin{table}
\caption{Comparison of CM success probability with filtering probability.}
\begin{tabular}{|l||d|d|d|d||}
\hline
$\gamma\bar t$	& 0.3	& 0.4	& 0.5	& 1.0	\\
\hline
\hline
$P_{seq,K=4}$   & 0.50  & 0.35  & 0.40  & 0.63  \\
\hline
${\rm Tr}_{_F}[\rho_{_F}(0)\rho_{_F}(\bar t)]$
                & 0.56  & 0.46  & 0.38  & 0.15  \\
\hline
\end{tabular}
\end{table}
%%%%%%%%%%%%%%%%%%%%%%%%%%%%%%%%%%%%%%%%%%%%%%%%%%%%%%%%%%%%%%
\end{document}